\documentclass[useAMS,usenatbib]{mn2e}
\usepackage[utf8]{inputenc}
\usepackage{graphicx}
\usepackage{multicol}
\usepackage{color}
\usepackage{tabularx}
\usepackage{longtable}

\usepackage{rotating}
\usepackage{pdflscape}
\usepackage{adjustbox}

\usepackage{graphics,epsfig,aas_macros,amsmath}

\newcommand{\nustar}{\textit{NuSTAR~}}

\newcommand{\rxte}{{\it RXTE~\/}}
\newcommand{\asat}{{\it AstroSat~\/}}
\newcommand{\bepposax}{{\it BeppoSAX~\/}}

\newcommand{\suzaku}{{\it Suzaku~\/}}

\newcommand{\code}{\texttt}
\usepackage{xcolor}
\usepackage{float}
\usepackage{subcaption}
\usepackage[colorlinks=true]{hyperref}

\usepackage[justification=centering]{caption}

\usepackage{rotating}

\usepackage{multirow}

\begin{document}

\title{Timing and spectral studies of Cen X-3 in multiple luminosity states using \asat}

\author[R. Bachhar et al.]
{Ritesh Bachhar$^{1,4}$, Gayathri Raman$^{1,2}$, Varun Bhalerao$^1$ and Dipankar Bhattacharya$^{3,5}$ \\ 
$^1$Indian Institute of Technology Bombay, Powai, Mumbai-400076, India\\
$^2$ Department of Astronomy and Astrophysics, The Pennsylvania State University, 525 Davey Lab, University Park, PA 16802, USA\\
$^3$ Inter University Center for Astronomy and Astrophysics (IUCAA), Post Bag 4, Ganeshkhind, Pune, Maharashtra-411007, India\\
$^4$ University of Rhode Island, South Kingstown, RI-02881, USA\\
$^5$ Ashoka University, Sonipat-131029, India}

\date{}

\maketitle

\begin{abstract}
We present the results of timing and spectral analysis of the HMXB pulsar, Cen X-3, with the help of observations carried out using the Large Area X-ray Proportional Counter (LAXPC) on board \textit{AstroSat}. As part of our analysis, we sampled the source properties during 4 different observation epochs covering two widely different intensity states. We obtain a timing solution and report precise measurements of the spin and orbital parameters corresponding to these observational epochs. The pulse profiles during the two intensity states reveal dramatically varying shapes within a time span of one month. We report the detection of one of the lowest measured frequencies of quasi periodic oscillations (QPO) at 0.026$\pm$0.001~Hz for Cen X-3 during its low intensity state. We also find correlated periodic and aperiodic noise components in the power density spectra. We further carried out a phase averaged and a pulse phase resolved spectral study, where we find that the best fit continuum spectrum is well described by an absorbed comptonization model along with a blackbody. Cen X-3 exhibited the presence of the $\sim$28~keV CRSF absorption line and a $\sim$6.6~keV Fe emission line in both the intensity states. Significant variations in the line forming regions and mode of accretion for Cen X-3 within time spans of a month make Cen X-3 a highly dynamic persistent binary. 

\end{abstract}


\section{Introduction}

High Mass X-ray Binaries (HMXBs) are some of the brightest X-ray emitters in the Milky Way galaxy. They comprise of a Neutron Star (NS) in orbit with a massive O or B type supergiant (see \citealt{PaulNaik2011} and references). A significant fraction of HMXBs harbour highly magnetized ($\sim$10$^{12}$ G) neutron stars which capture stellar winds from the companion star. In rare cases, HMXB pulsars are observed to show signatures of disk accretion from a Roche lobe filling companion. Inside the NS's strong magnetosphere, the accreted matter falls on to the magnetic poles of the pulsar through an accretion column, while releasing X-rays (see \citealt{kretschmar2020} for a recent review). The observed characteristic pulsations are a result of the misalignment between the rotation and magnetic axes. Factors that contribute to the pulse profile variations mainly include the accretion rate, magnetic field geometry, beaming patterns, geometric configuration of the accretion column and the pulsar's orientation \citep{Nagase1989}. Many transient and persistent HMXB pulsars also exhibit variations in their pulse profiles, centroid line energy of the Cyclotron Resonant Scattering Feature (CRSF) and even Quasi Periodic Oscillations (QPOs) (see \citealt{Caballero2012MmSAIreview,PaulNaik2011,Kretschmar2019} for reviews of X-ray pulsars and their properties). Detailed studies of these characteristics as a function of their luminosity state have been undertaken in very few cases \citep{Becker2012,Mushtukov2015,Kretschmar2019}. For this work, we focus on these state dependent properties in one such well profiled persistent galactic X-ray pulsars, Cen X-3.

Cen X-3 was the first ever discovered HMXB pulsar \citep{Chodil1967,Giacconi1971}. Using follow-up Uhuru observations, Cen X-3 was found to be an eclipsing pulsar binary with an orbital period of $\sim$2.1 d that exhibited periodic pulsations every 4.8 s \citep{Schreier1972}. The optical companion was first identified as a giant O type star with a source distance of $\sim$8~kpc \citet{Krz1974}. Using radial velocity measurements from optical spectra, the neutron star mass was measured to be 1.21$\pm$0.21 M$_{\odot}$ and the optical companion of Cen X-3 was determined to be an O 6-8 III supergiant star (V779 Cen) with a mass of 20.5$\pm$0.7 M$_{\odot}$ \citep{Hutchings1979,Ash1999}. More recently, \citet{Thompson-Rot2009} estimated the distance to the source as 5.7$\pm$1.5~kpc using energy-resolved dust scattered X-ray halo studies using Chandra observations. Latest Gaia measurements indicate a distance of 8.7~kpc \citep{gaia}. For the sake of this work we have considered the more reliable Gaia measurement of 8.7$\pm$2.3~kpc obtained by parallax inversion. The high source X-ray luminosity (10$^{37}$~ergs~s$^{-1}$, \citealt{Suchy2008}), optical light curve \citep{vanP1983} and the presence of QPOs \citep{Takeshima1991,Raichur2008} in Cen X-3 are all strong factors that suggests that the dominant mode of accretion is via an accretion disk fed by a companion, probably at the onset of atmospheric Roche Lobe overflow.



For Cen X-3, the broadband continuum spectrum (0.1--100~keV) has been more or less effectively characterised using an absorbed power law with a high energy cutoff at energies between 10.0--20.0~keV \citep{tomar2021}. Soft X-ray excess below 1~keV was accounted for using blackbody models with kT$\sim$0.1~keV \citep{Burderi2000}. Several studies have also reported the detection of Fe-K emission lines  at 6.4~keV, 6.67~keV and 6.97 keV \citep{Nagase1989,tugay2009,Ebisawa1996,PaulNaik2011}. The presence of a CRSF at $\sim$28~keV was first reported by \citet{Santangelo1998}, which implied a NS surface magnetic field of $\sim$ 3$\times$10$^{12}$ G. Several more CRSF measurements have been reported (see \citealt{tomar2021} for a compilation of all measurements). Cen X-3 has also exhibited a range of QPOs in the mHz (40--90~mHz: \citealt{Raichur2008}) as well as in the kHz regime \citep{Garrett1999}.

In this paper we explore the luminosity state dependent timing and spectral properties of this disk accreting pulsar system using publicly available \asat observations carried out in late 2016 and early 2017. The paper is organized as follows: The observations and data reduction methods are outlined in Section 2, followed by the timing and spectroscopy results in Sections 3.1 and 3.2, respectively. We then discuss the implications of our long term study in Section 4 and summarize our work in Section 5. 

\section{Observations and data reduction}
\asat is India's first multi-wavelength astronomy satellite launched in September, 2015. \asat consists of five different on board payloads spanning a wide electromagnetic spectral range from the optical/UV all the way up to hard X-rays \citep{Singh2014}. The Large Area X-ray Proportional Counter (LAXPC) instrument on board \asat comprises of 3 identical, co-aligned detectors (LXP10, LXP20 and LXP30) with a field of view of 1$^\circ \times$1$^\circ$. LAXPC has an effective area of about 6000 cm$^2$  and an  absolute timing resolution of 10 $\mu$s in the 3.0--80.0~keV energy range. Detailed description of the instrument and calibration can be found in \citet{Yadav2017}. Owing to its high time resolution and broad band sensitivity, LAXPC is best suited to carry out spectro-timing studies of X-ray pulsars.  The Soft X-ray Telescope (SXT) consists of a focusing telescope with a CCD detector. With an effective area of $\sim$90 cm$^2$, it is capable of performing X-ray imaging and medium resolution spectroscopy in the 0.3--8.0~keV energy range \citep{Singh2017}. Cen X-3 was observed using LAXPC and SXT simultaneously between Dec 2016 - Feb 2017 as part of the Announcement of Opportunity observing cycle. We refer to the 4 data sets analyzed in this work as A, B, C and D. Details of the Obs-IDs and exposures are specified in Table \ref{tab:Obs table}.  
The LAXPC observations were carried out in the Event Analysis (EA) mode. The Level 1 data from all the 4 data sets were reduced using the LAXPC Data analysis pipeline version 3.1\footnote{Data analysis software was obtained from http://astrosat-ssc.iucaa.in/?q=laxpcData}. The pipeline code combines data from multiple orbits and also filters out overlapping segments between each orbit. Using the standard pipeline \code{laxpc\_make\_event}, we then generate the event file. A good time interval (GTI) window was applied during the processing in order to exclude the time intervals corresponding to the Earth occultation periods, SAA passage and standard elevation angle screening criteria. This was created using the \code{laxpc\_make\_stdgti} tool. During the years 2016--2017, the gain of the LXP30 detector varied unpredictably due to gas leakage. We, therefore, do not use LXP30 observations for any analysis. Further, since the spectral response for the LXP20 detector unit was not reliable during these observations, we discard spectral analysis studies using this detector and use only the LXP10 detector unit. All spectral products for this LXP10 detector were extracted using the task \code{laxpc$\_$make$\_$spectra}.

The SXT observations were carried out in the Photon Counting (PC) mode, where the CCD camera takes successive 2.4~s exposures. The LEVEL 1 data were processed using the SXT software pipeline version 1.4b and the SXT spectral redistribution matrices in CALDB (v20160510).  Orbit-wise cleaned event files were merged and updated with the corrected exposure time using the SXT merger tool called \code{sxt\_merger\_make}. We selected all the events inside a circular region of radius $\sim$10$^\prime$, centered on the source co-ordinates. The image, light curves and spectra were extracted using Xselect v 2.4 in Heasoft 6.26. We identified jitter in the SXT images corresponding to Obs IDs B, C and D, which is why we decided against including SXT for further analysis in this paper.

\begin{table*}
    \centering
    \begin{tabular}{|c|c|c|c|c|c|}
    \hline
      Observation ID    & Start Time   & Exposure  & Average LAXPC  & Average SXT & Average Swift-BAT \\
      &  (UTC) & (ks)  & count rate (c/s) & count rate$^{a}$ (c/s) & count rate (c/s)\\
       
       \hline
      Obs A - G06$\_$091T01$\_$9000000880  &  2016-12-12 12:31:55.7 &  39.6 & 3170 & 7.2 &0.058\\
      Obs B - A02$\_$111T01$\_$9000000954  & 2016-12-19 11:21:56 & 15.0 & 3029 & 5.4 &0.054 \\
      Obs C - A02$\_$111T01$\_$9000000954 &  2017-01-09 16:43:01.8 & 18.5 & 1432 & 3.5 & 0.02\\
      Obs D - A02$\_$111T01$\_$9000000986  &  2017-01-27 16:30:43.2 &  17.8 & 1534 & 4.8 & 0.03 \\
         \hline
    \end{tabular}
    \caption{Details of LAXPC observations of Cen X-3 used for this work. The trend seen in the average 3.0--80.0 keV LAXPC count rates reflect the trend observed in the count rate average as measured by 15.0--50.0~keV Swift-BAT monitor.\\
    $^{a}$ Average SXT count rate has been estimated excluding the sudden data drops.}
    \label{tab:Obs table}
\end{table*}

       


\section{Analysis and results}
\begin{figure}
    \centering
    \includegraphics[scale=0.5,angle=0]{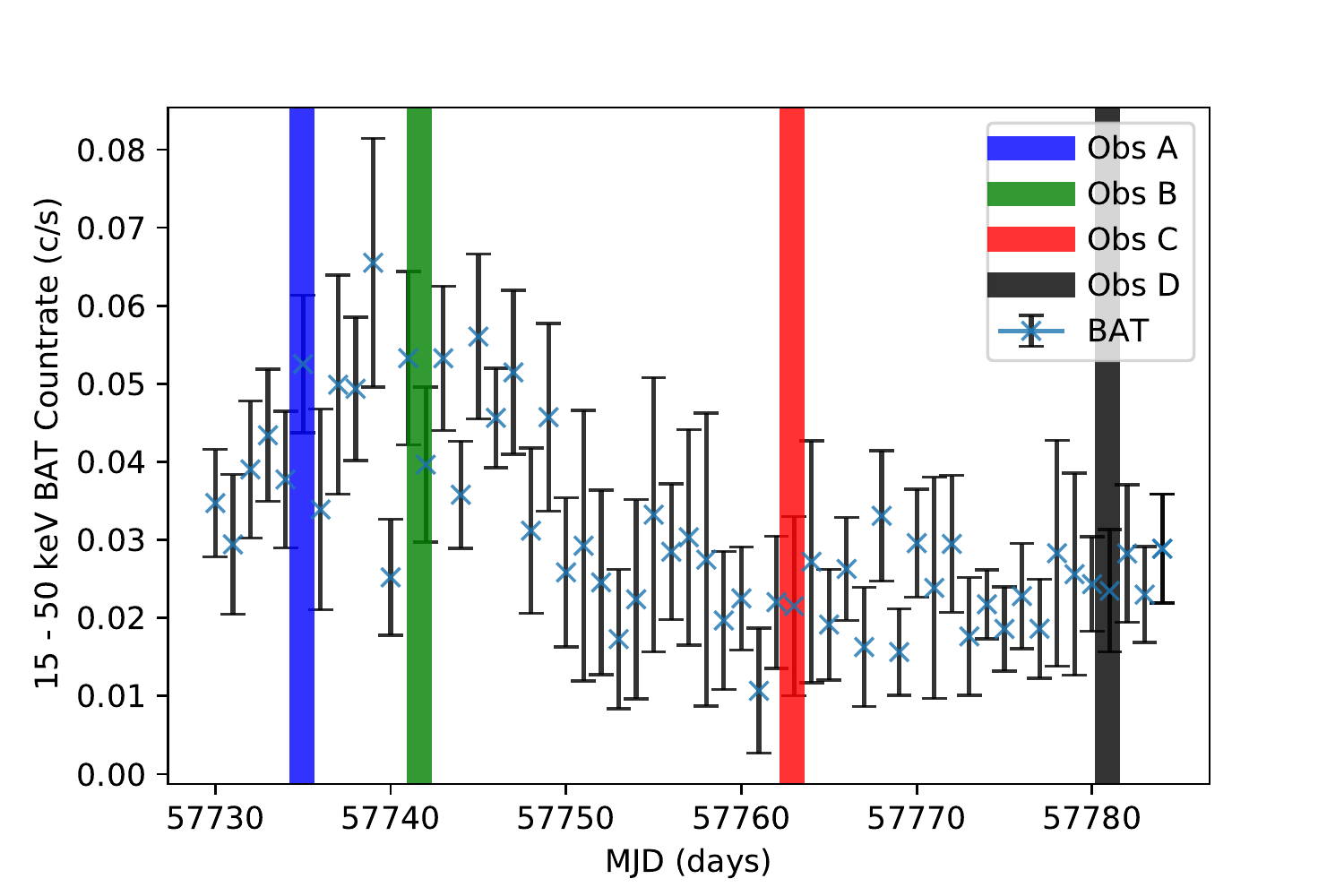}
    \caption{The 15--150~keV Swift/BAT light curve of Cen X-3 indicating the location of each of the four \asat observations. As the figure indicates, Obs A (blue) and Obs B (green) correspond to a high source intensity state, while Obs C (red) and D (black) correspond to a low intensity state. The source intensity states were identified based on the average source count rate as well as the source fluxes obtained from spectroscopy (see Section 3.2). }
    \label{fig:bat-lc}
\end{figure}

 We obtained the long term 15--150~keV Swift/BAT \citep{Barthelmy2005} light curve of Cen X-3 to identify the intensity state during which the \asat observations were carried out. Two of the observations, Obs A and Obs B were obtained during a high intensity state whereas Obs C and Obs D correspond to a low intensity state. Figure \ref{fig:bat-lc} shows the location of these 4 \asat observations marked on the Swift/BAT light curve. The 15--50~keV Swift-BAT count rate at the time of the \asat observations is shown in Table~ \ref{tab:Obs table}. 

A uniform timing analysis procedure was adopted for all four observations mentioned in Table \ref{tab:Obs table}. We extract the light curve and study the pulse profiles and power density spectra, details of which are elaborated in the following subsections. 

\subsection{Timing analysis} 

The source and the background light curves were extracted using the tasks \code{laxpc\_make\_lightcurve} and \code{laxpc\_make\_backlightcurve}, respectively, with a time bin size of 10 ms. The background was subtracted using the \code{FTOOLS} task \code{lcmath}. We carried out barycenter correction using the \code{as1bary} tool. The average source count in all the four Obs-IDs is indicated in Table \ref{tab:Obs table}. The LAXPC source count rate showed a steady modulation around a mean of $\sim$3000 c/s for Obs A and Obs B, while the mean count rate decreased to $\sim$1500 c/s for Obs C \& Obs D. No eclipse is seen during the observations. 
\begin{figure}
    \centering

    \includegraphics[scale=0.43,angle=-90,trim={0 0.5cm 0cm 2.1cm},clip]{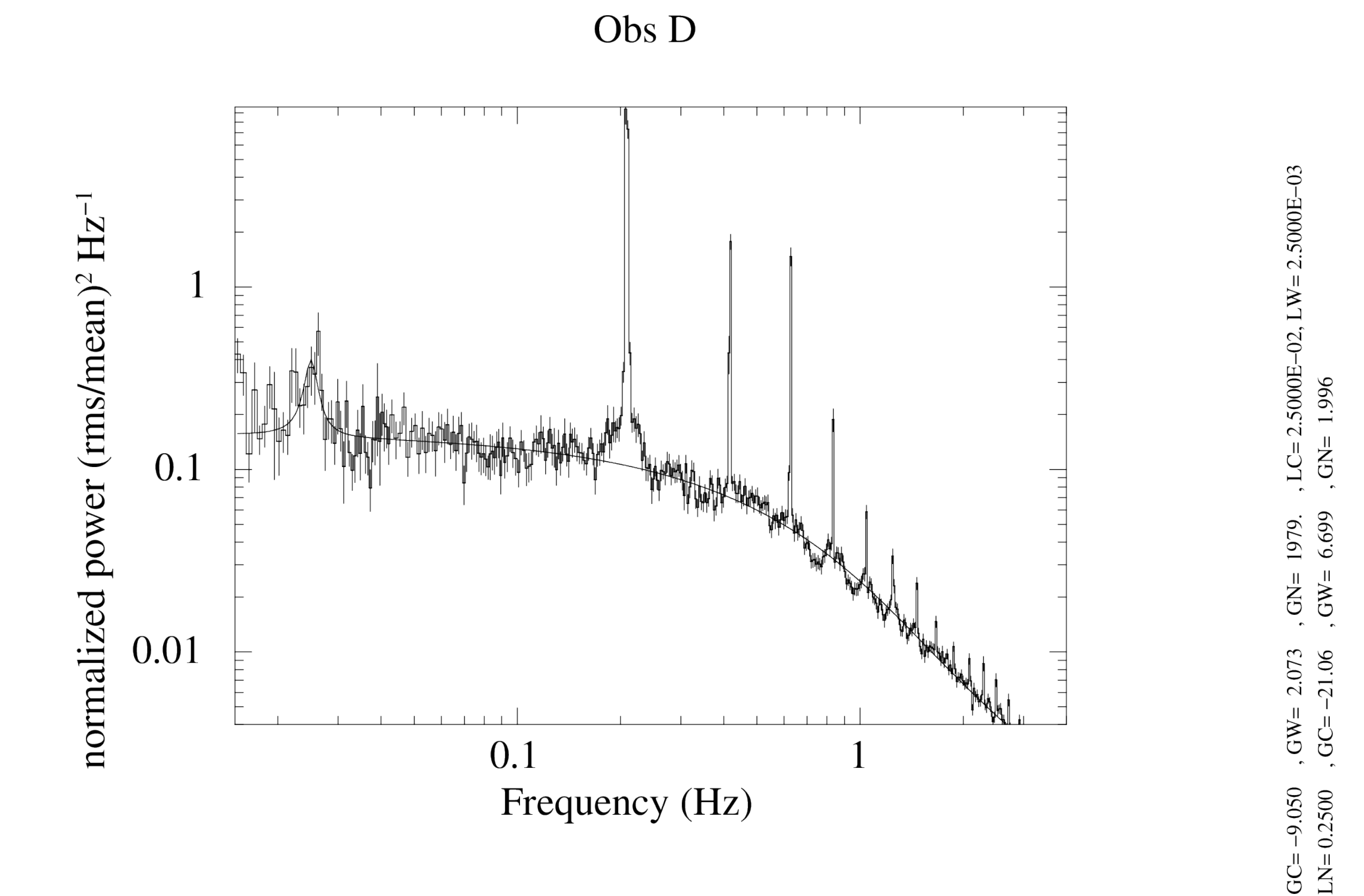}
    \caption{Figure shows the power density spectrum for Obs ID D showing the pulse peak at 0.12~Hz and several of the harmonics. The RMS normalised power drops considerably beyond 10~Hz. We detect a $\sim$26~mHz QPO in the lower luminosity state observation D, which has been modeled using a Lorenztian function (solid black curve).}
   \label{fig:pds}
\end{figure}

We generated the power density spectrum (PDS) for the four observations of Cen X-3 to identify pulsations and search for QPOs. We carried out a fourier transform of the 10 ms light curve from all the LAXPC detectos using the \code{FTOOLS} task -- \code{powspec}. All $\sim$200 interval segments, each with 8192 bins, were averaged in order to generate the final PDS. The power spectrum was subtracted for white noise and normalised such that the integral gives the square of RMS fractional variability. The PDS for the Obs ID D is shown in Figure \ref{fig:pds}. 

The 0.01--100~Hz PDS shows a strong pulsation peak at ${\sim}0.2$~Hz, corresponding to the NS spin frequency along with multiple harmonics with decreasing peak amplitude.  Interestingly, the main pulse peak and the harmonics show a broadened base, which is a feature reported for several other pulsars like GX 304-1 \citep{Jincy2011}, 4U 1901+03 \citep{Marykutty2011}, etc, as well. Such wings in the HMXB pulsars could arise due to the presence of correlated periodic and aperiodic noise components in the PDS \citep{Burderi1993,LazztiStella1997}. We searched for QPOs near the previously reported mHz \citep{Raichur2008} and kHz \citep{Jernigan1999} frequency ranges. We detected a QPO feature in only one of the four observations - Obs ID D. We fit the PDS in the frequency range 0.01--0.04~Hz with a broad gaussian to account for the continuum around the QPO. Chi-square statistics have been used to perform the fit in QDP/PLT. We then introduce a lorentizan to describe the broad QPO feature centered around 26~mHz. We obtain a best fit for the QPO at 0.026$\pm$0.001 Hz with a Q factor (=$\frac{\nu}{\Delta \nu}$) of 10.2. The uncertainties are quoted at 90\% level. The RMS normalized power for the QPO is 3.1\%$\pm$0.1\%. We also examined the strength of the QPO feature in different energy bands (same energy bands as used to examine the pulse profiles). The measured rms normalized power, has values -  3.7\%$\pm$0.1\%, 2.4\%$\pm$0.1\%, 2.4\%$\pm$0.1\%, 2.4\%$\pm$0.1\% and 4.6\%$\pm$0.1\% in the 3.0--6.0~keV, 6.0--10.0~keV, 10.0--15.0~keV, 15.0--22.0~keV and 22.0--35.0~keV bands, respectively. Interestingly, we find a tentative strengthening of the QPO RMS in the 22.0--35.0~keV energy band which brackets the CRSF line ($\sim$28~keV). However, due to limited statistics  in some of the energy bands, our fits were unable to constrain the width and strength of the QPO. We therefore cannot robustly confirm this behavior using these observations.

\begin{figure*}
\centering
    \includegraphics[scale=0.45]{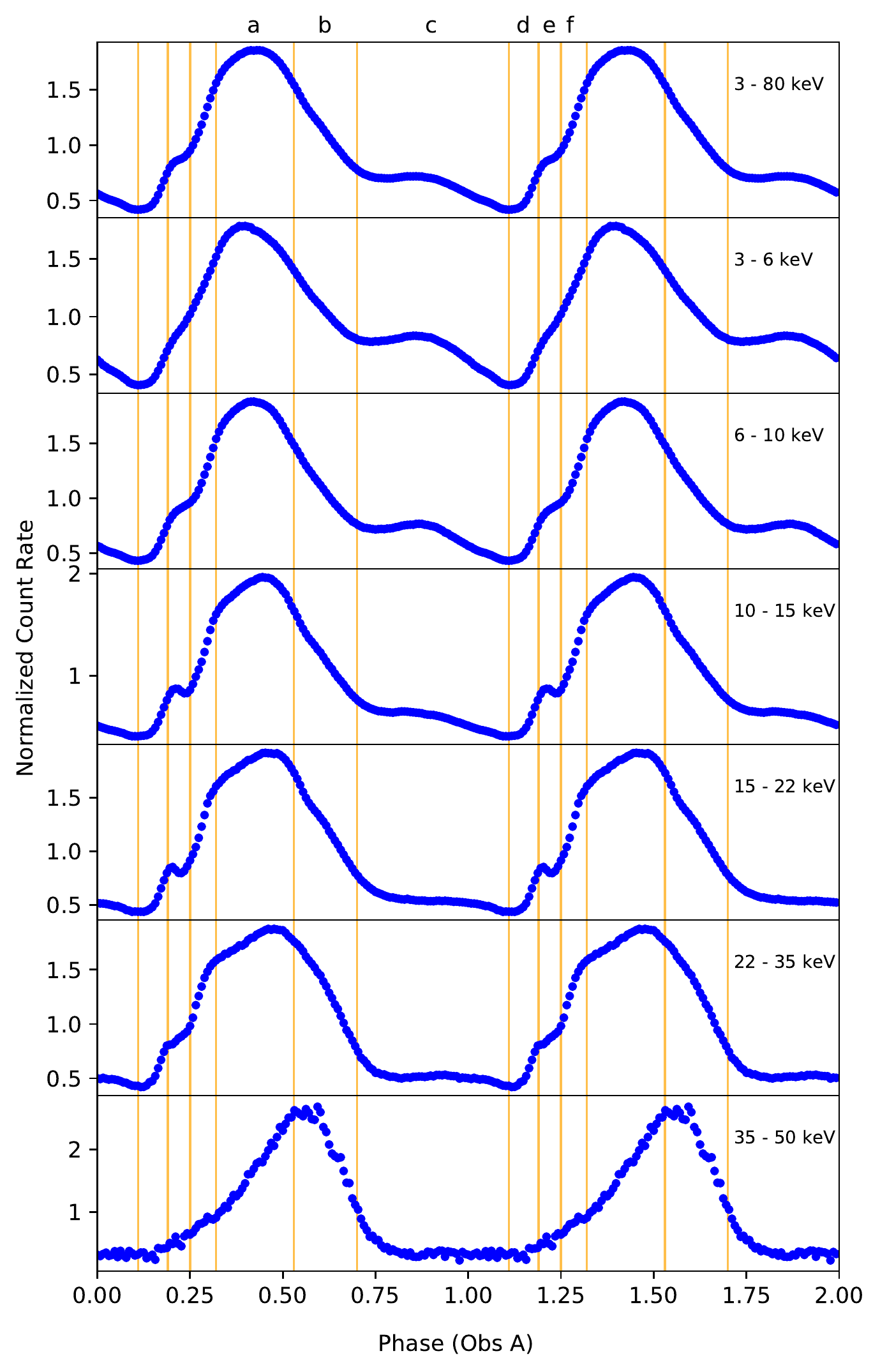}
    \includegraphics[scale=0.45]{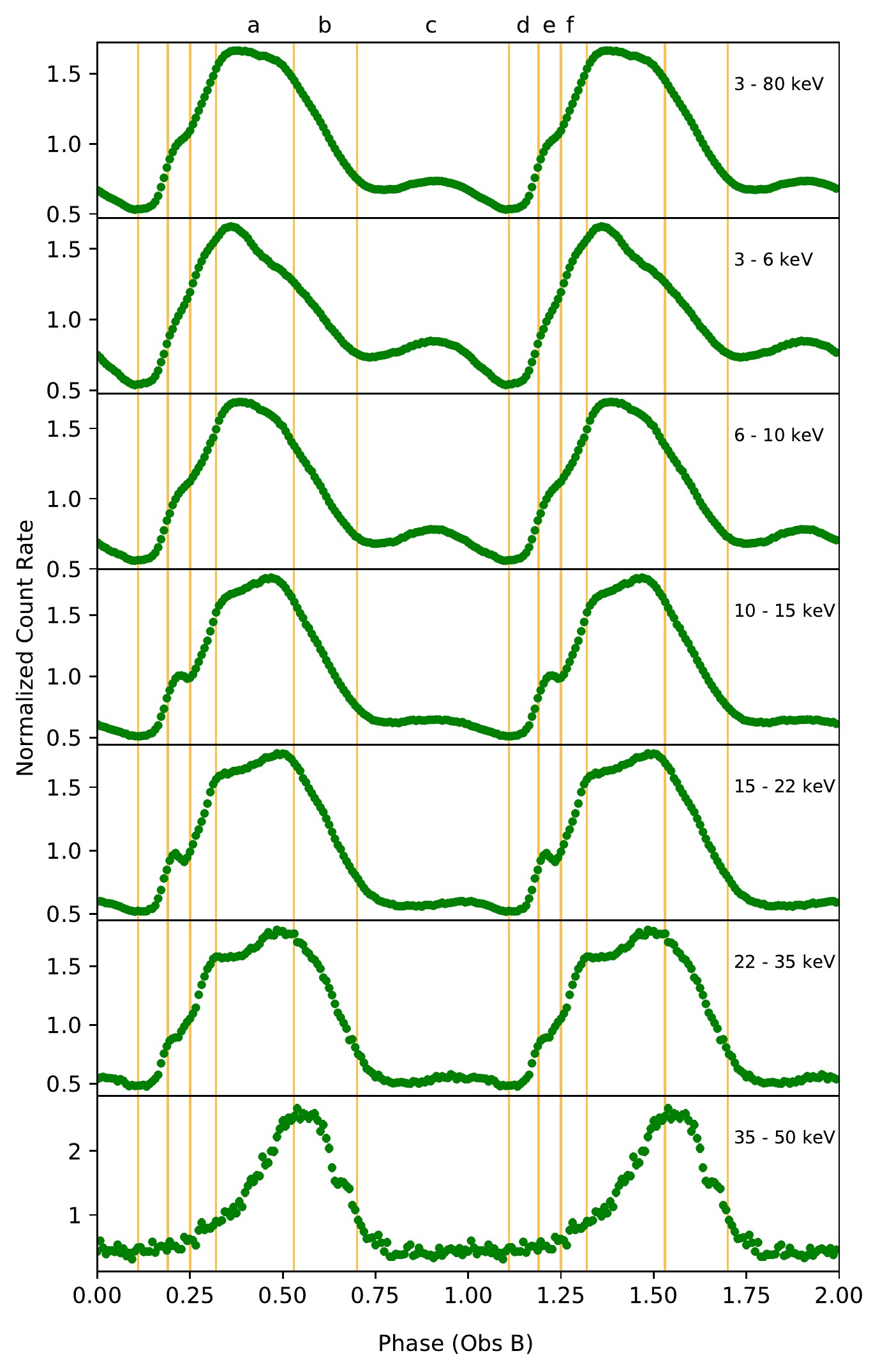}
    \includegraphics[scale=0.45]{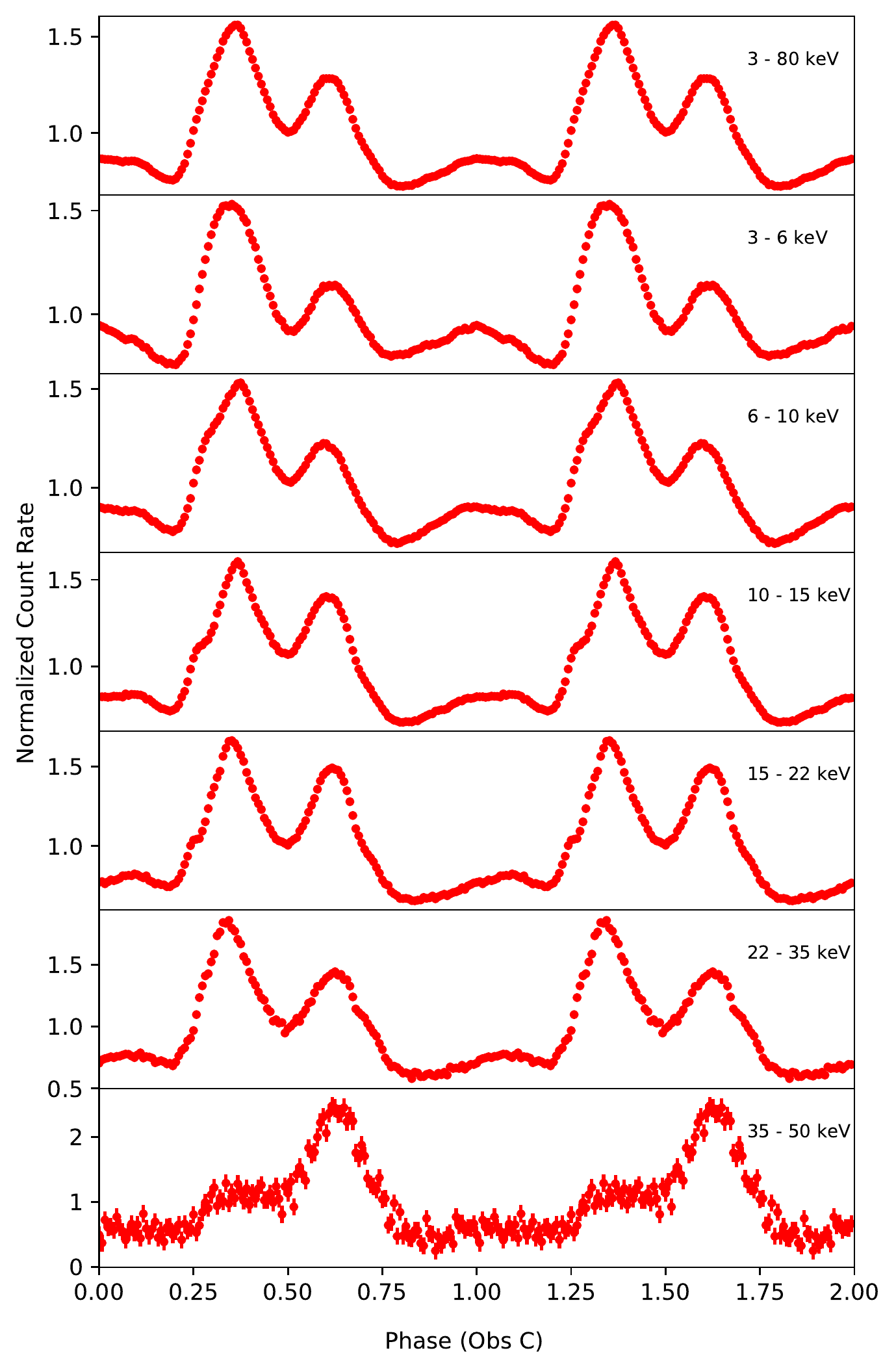}
    \includegraphics[scale=0.45]{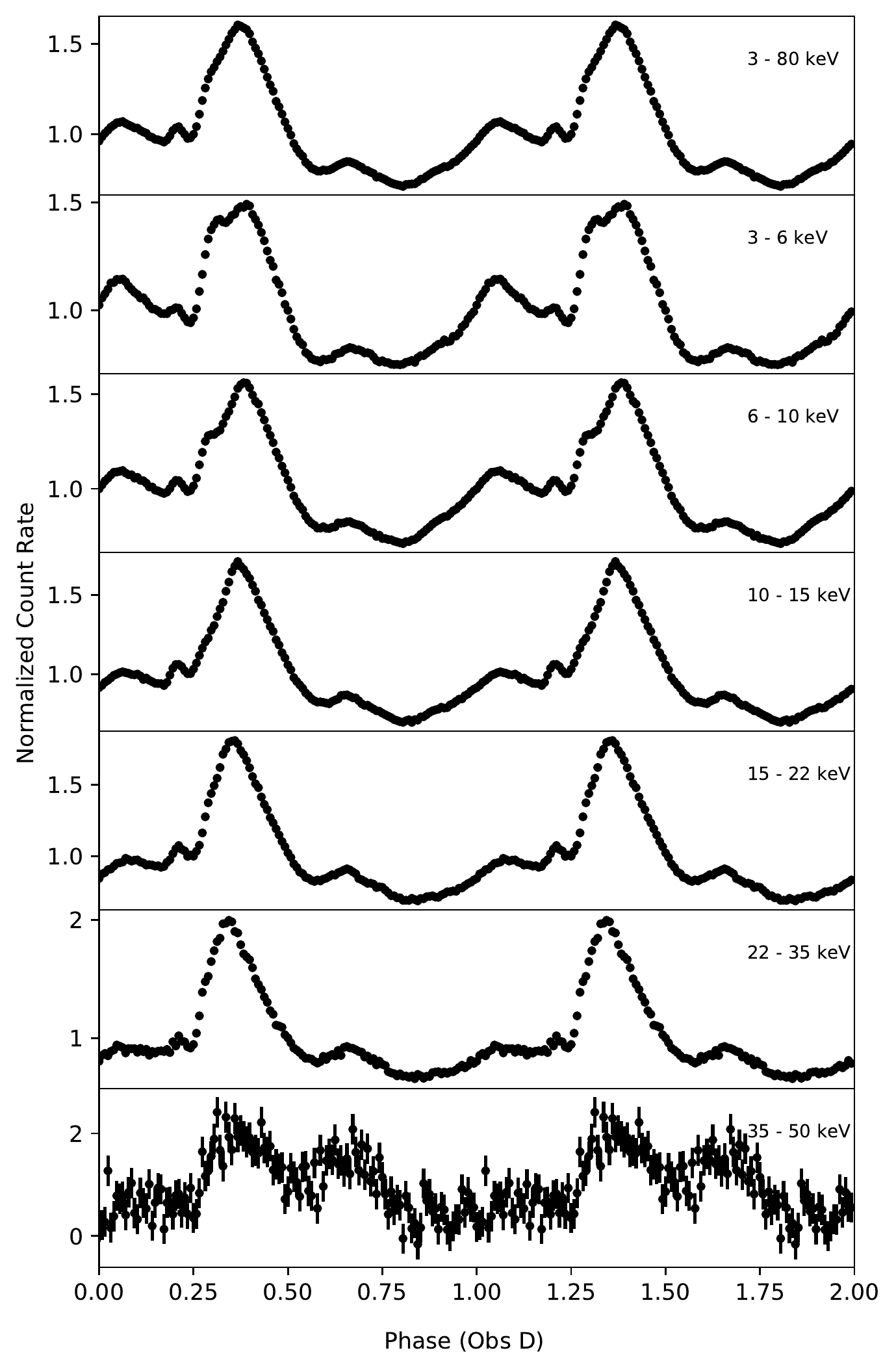}
    \caption{Energy resolved pulse profile from all four observations, Obs A (Upper left-blue), Obs B (Upper right-green), Obs C (Lower left-red), and Obs D (Lower right-black) were folded at a pulse period of 4.80183~s, 4.80196~s, 4.80170~s, and 4.80323~s. The pulse profiles for the different intensity state observations show distinctly different shapes. Each panel corresponds to different energy range and normalized pulse profile is plotted for two consecutive phases. The vertical yellow lines in Obs A and Obs B represent phase boundaries used for phase resolved analysis (Section \ref{section:pulse_phase}). The data for the fainter observations (Obs C and Obs D) do not have high enough signal-to-noise ratio for phase resolved analysis.}
    \label{fig:pulseprofiles}
\end{figure*}

\begin{figure}
    \centering
\includegraphics[scale=0.4]{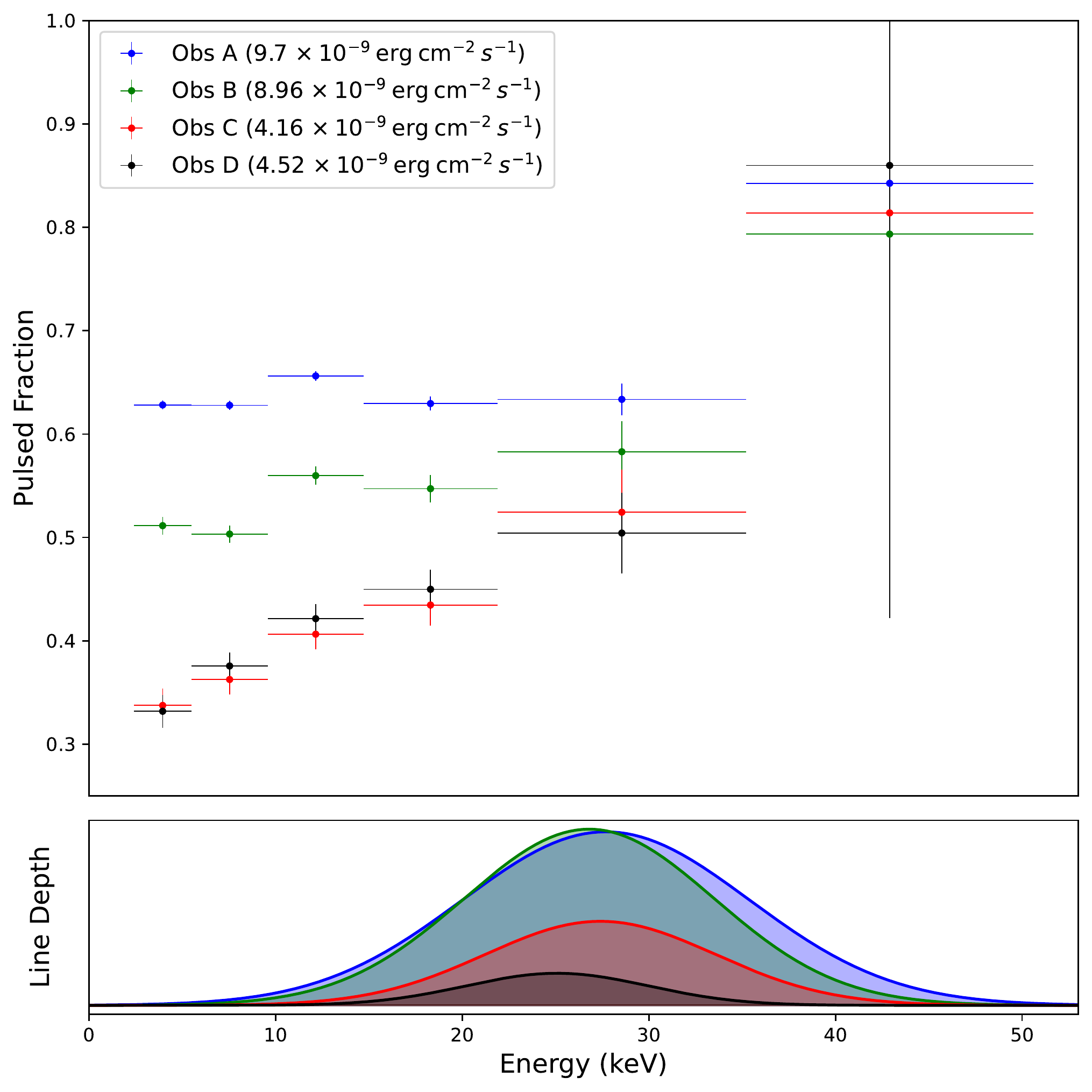}

\caption{Pulsed fraction for the 3.0--80.0~keV folded pulse profiles for all the Obs IDs plotted as a function of energy for Cen X-3. In the faint epochs (Obs C \& Obs D) we see a strong trend of increasing pulsed fraction with energy, while the variation is less pronounced for the bright intensity state epochs (Obs A \& Obs B). Bottom panel indicates the CRSF depth with the Gaussian centre at the line energy with the spread representing the line width. }
    \label{fig:PFrac}
\end{figure}

In order to constrain the pulsar's spin period with high precision in the four Obs IDs, we carried out a sensitive period search analysis. The arrival time of the photons was corrected to the solar system barycenter. Using existing ephemeris \citep{falanga2015, raichur2009, parisree}, we corrected for the the photon arrival time delay due to the orbital motion and extracted sensitive period measurements from the resulting time series for each observation. Then we folded the 10ms binned light curves using the \code{FTOOL} task \code{efold} at the intrinsic spin period of each observation. The average pulse profiles in the 3.0--80.0~keV energy band are shown in Figure \ref{fig:pulseprofiles}. Once again, the two intensity states show clearly distinct pulse profile shapes. The high intensity observations (Obs A \& Obs B) show a profile with a single narrow peak at phase $\sim$0.5 and a smaller amplitude `knee' bump at phase $\sim$0.85. Within the lower intensity observations, Obs C pulse profile shows a strong double peaked profile with peaks at phases 0.3 and 0.6, whereas in the Obs D, the pulse profile shows 3 narrow peaks at phases 0.2, 0.4 and 0.7. In order to further examine the variations in the pulse profiles, we generated energy resolved pulse profiles in the energy bands 3.0--6.0~keV, 6.0--10.0~keV, 10.0--15.0~keV, 15.0--22.0~keV and 22.0--40.0~keV. We estimate the pulsed fraction in the various energy bands using the prescription (I$_{\rm max}$--I$_{\rm min}$)/(I$_{\rm max}$+I$_{\rm min})$. We observe that the pulsed fraction increases as a function of energy for Obs C \& D as shown in Figure \ref{fig:PFrac}. At energies below 30~keV, the trend is weak for Obs A \& B. We also do not observe any local enhancements of the pulsed fraction around the CRSF energy.

\subsection{Spectroscopy}

\subsubsection{Phase averaged spectroscopy}

We extracted LAXPC spectra from all the detector layers. Since LAXPC has a poor response below $\sim$5~keV, we kept the hydrogen column density ($\mathrm{N_H}$)  fixed to the galactic absorption value of $1.11\times\,10^{22}$ ~atoms cm$^{-2}$ \citep{GalaxySurvey} throughout our analysis\footnote{\url{https://heasarc.gsfc.nasa.gov/cgi-bin/Tools/w3nh/w3nh.pl}}. Several previous works on broadband spectral analysis of Cen X-3, have used a simple power law \citep{PaulNaik2012BASI}, the combination of a power law and a blackbody \citep{Ferr2021}, a power law with a high energy cutoff alongside a blackbody \citep{Burderi2000}, or in some cases a power law model modulated by partial covering \citep{NaikPaulAli2011}, \citep{Torregrosa}. \citet{Farinelli2016} discuss various difficulties while adopting more complicated models such as \code{compmag} (a hybrid model that was developed to describe accretion onto a magnetized NS, \citealt{FarinelliModel2012}) for fitting broadband Cen X-3 \suzaku, \bepposax and \nustar data.

In our current work, in order to describe the 3.0--50.0~keV LXP10 continuum spectra, we started with a few standard XSPEC models like i) a simple \code{powerlaw} model along with a \code{blackbody} and ii) a \code{powerlaw} with \code{highEcut} \citep{White1983}. Our spectral results reflect the fact that a powerlaw modulated by a highEcut does indeed offer a reasonable fit, without the need for a blackbody or any partial covering. Previous works such as \citet{Burderi2000} have always required an additional blackbody component while using the cutoff power law model. While trying to export this simple phenomenological model to the pulse phase resolved spectra (Section 3.2.2), we were unable to constrain any of the parameter values. This prompted us to explore other alternative comptonization models as has been recently adopted for some pulsars like IGR 19294+1816, etc. \citep{Tsygankov2019,Raman2021}. We then proceeded to carry out spectral fits using iii) the \code{compTT} model \citep{Titarchuk1994} and iv) the \code{nthcomp} model. The \code{nthcomp} model describes the thermal comptonization of soft photons in a hot plasma \citep{Zd1996,Z1999}. Reasonable fits were obtained using the \code{nthcomp} model, while the \code{compTT} model proved to be a poor choice. Since the \code{nthcomp} model (Model I, from now) showed a residual at low energies, we added a black-body component which improved our overall continuum fit. Moreover, since the blackbody model was sufficient for the fit, we did not additionally consider any partial covering for this work. We further assume that blackbody photons from the NS surface are comptonized, hence we link the blackbody temperature and the seed photon temperature parameter of the compton emission. We note that \code{powerlaw} with \code{highecut} (Model II, from now) fits become worse with addition of the blackbody component, and provides inconsistent fits in pulse phase resolved analysis (as we describe in Section 3.2.2), while on the other hand, Model I not only constrains the pulse phase averaged spectral parameters better, but also translates smoothly into the pulse phase resolved analysis. Our spectral fits indicate that Model I (\code{nthcomp+bbody}) is statistically preferred over Model II (\code{powerlaw$\times$highEcut}).

\begin{figure}
    \centering
    \includegraphics[scale=0.45]{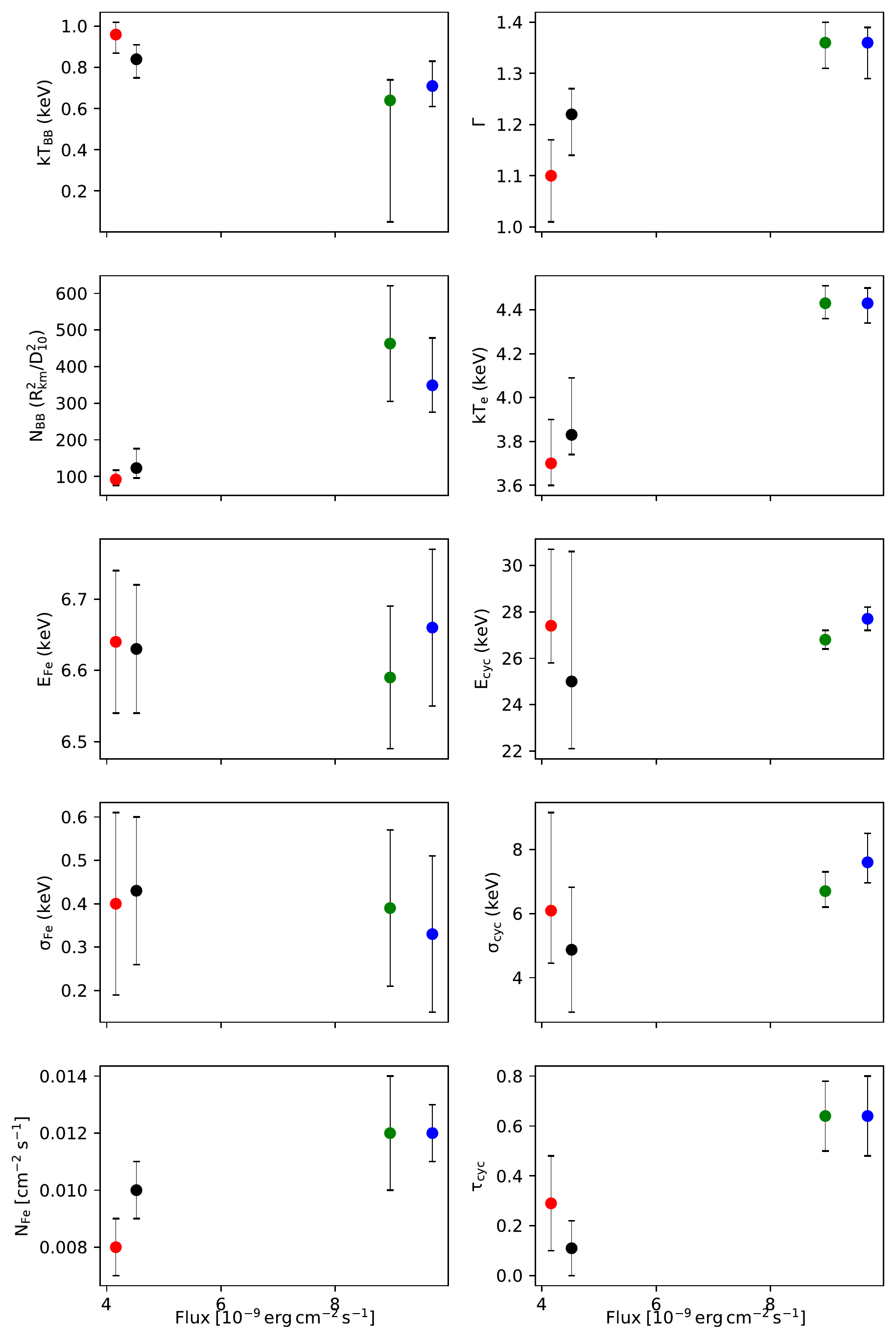}
    \caption{Variation of phase-averaged spectral parameters with intensity state of the source. Blue, green, red and black represent Obs A, Obs B, Obs C, and Obs D, respectively. The various model components are as in Table \ref{tab:phase_avg}.}
    \label{fig:flux-vs-param}
\end{figure}

We observed narrow residuals near $\sim$28~keV  which we modeled using a gaussian absorption line. We observe that the CRSF absorption is strongest in the high luminosity states. Residuals at $\sim$6.6~keV were fit by adopting a broad gaussian emission feature ($\sigma_{Fe}\sim$0.4~keV). We allowed for a 1.5\% systematics while fitting the spectra. The details of the best fit spectra parameters for both models are shown in Table \ref{tab:phase_avg} and the variation of the spectral parameters in all four intensity states is indicated in Figure \ref{fig:flux-vs-param}.

 We further estimate the source flux (for the Model I case) in the 3.0--50.0~keV energy range for all four Obs IDs as  $9.7 \times 10 ^{-9}$~erg~cm$^{-2}$~s$^{-1}$, $8.96 \times 10^{-9}$~erg~cm$^{-2}$~s$^{-1}$, $4.2 \times 10^{-9}$~erg~cm$^{-2}$~s$^{-1}$ and $4.6 \times 10^{-9}$~erg~cm$^{-2}$~s$^{-1}$, respectively. Assuming a source distance of 8.7~kpc \citep{Treuz2018}, we obtain the X-ray luminosities as 
 $9.0 \times 10^{37}$, $8.2 \times 10^{37}$, $3.8 \times 10^{37}$ and $4.2 \times 10^{37}$ erg~s$^{-1}$ for the four Obs IDs, respectively.



\begin{figure}
    \centering

    \includegraphics[scale=0.35,trim={1cm 1cm 0 1.3cm},clip]{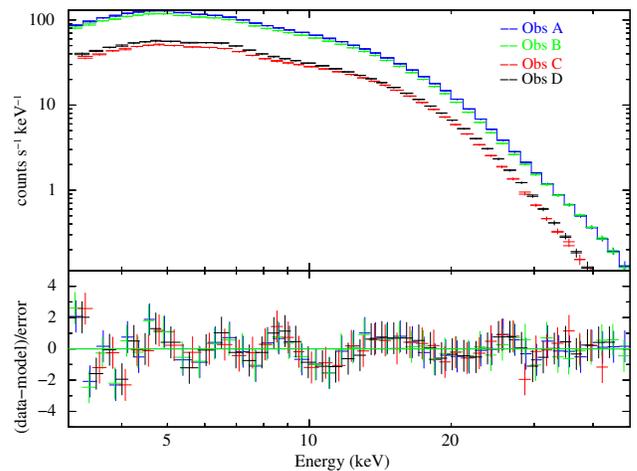}

    \caption{Best fit LAXPC spectra accumulated over all the four observations. These phase averaged spectra are fit with comptonized continuunm, a blackbody component and a Gaussian line mimicking the Fe emission line near 6~keV. Fit residuals are shown in the lower panel. }
    \label{fig:phase_avg_spec}
\end{figure}

\begin{table*}
    \centering
     \caption{Best fit spectral parameters for phase averaged spectral analysis of the LXP10 detector for all four Obs IDs. Model I and Model II are as described in Section 3.2. Errors quoted are for 90\% confidence range. \\ $^a$ photons keV$^{-1}$~cm$^{-2}$~s$^{-1}$ at 1~keV.\\ $^b$ Flux is in units of $10^{-9}$ ergs$\; \rm{cm}^{-2}\rm s^{-1}$ and is in 90\% confidence range. }
    \label{tab:phase_avg}
    \begin{sideways}
    \begin{tabularx}{\textheight}{|p{1.1cm}|p{2.3cm}|c|c|c|c|c|c|c|c|} 
    \cline{1-10}
      \multirow{2}{*}{Model } & \multirow{2}{*}{Parameters} & \multicolumn{2}{c|}{Obs A} & \multicolumn{2}{c|}{Obs B} & \multicolumn{2}{c|}{Obs C} & \multicolumn{2}{c|}{Obs D} \\[2pt]
      \cline{3-10}
      & & Model I & Model II & Model I & Model II & Model I & Model II & Model I & Model II \rule{0pt}{3ex}\\[2pt]
      \cline{1-10}
      & & & & & & & & & \\
      \code{TBabs} & nH ($10^{22}\;$cm$^{-2}$) & $1.11$ (fixed) & $1.11$ (fixed) &  $1.11$ (fixed) & $1.11$ (fixed) &  $1.11$ (fixed) & $1.11$ (fixed) &  $1.11$ (fixed) & $1.11$ (fixed) \\ [5pt]
      \cline{1-10}
      & & & & & & & & & \\
      \code{nthcomp} & $\Gamma$ & $1.36^{+0.03}_{-0.07}$ & -- & $1.36^{+0.04}_{-0.05}$ & -- & $1.10^{+0.07}_{-0.09}$ & -- & $1.22^{+0.05}_{-0.08}$ & -- \\ [5pt]
      & kT$_{\rm e}$ (keV) & $4.43^{+0.07}_{-0.09}$ & -- & $4.43^{+0.08}_{-0.07}$ & -- & $3.7^{+0.2}_{-0.1}$ & -- & $3.83^{+0.26}_{-0.09}$ & -- \\ [5pt]
      & norm & $0.13\pm0.05$ & -- & $0.14\pm0.04$ & -- & $0.009\pm0.004$ & -- & $0.027^{+0.006}_{-0.01}$ & -- \\ [5pt]
      \code{blackbody} & kT (keV) & $0.71^{+0.12}_{-0.10}$ & -- & $0.64^{+0.10}_{-0.59}$ & -- &  $0.96^{+0.06}_{-0.09}$ & -- & $0.84^{+0.07}_{-0.09}$ & -- \\ [5pt]
      & norm & $349^{+129}_{-73}$ & -- & $463\pm1582$ & -- & $92^{+25}_{-16}$ & -- & $123^{+53}_{-27}$ & -- \\ [5pt]
      \cline{1-10}
      & & & & & & & & & \\
      \code{powerlaw} & PhoIndex & -- & $1.24\pm0.02$ & -- & $1.24\pm0.02$ & -- & $1.16\pm0.01$ & -- & $1.21\pm0.01$ \\[5pt]
      & norm$^a$ & -- & $0.59\pm0.02$ & -- & $0.54\pm0.02$ & -- & $0.203\pm0.003$ & -- & $0.24\pm0.06$ \\[5pt]
      \code{highEcut} & cutoffE (keV) & -- & $13.2\pm0.2$ & -- & $13.1\pm0.2$ & -- & $14.6\pm0.1$ & -- & $14.6\pm0.1$ \\[5pt]
      & foldE (keV) & -- & $8.7^{+0.2}_{-0.1}$ & -- & $8.9\pm0.1$ & -- & $7.5\pm0.1$ & -- & $7.7\pm0.1$ \\[5pt]
      \cline{1-10}
      & & & & & & & & & \\
      \code{gaussian} & $\mathrm E_{\mathrm Fe}$ (keV) & $6.6\pm0.1$ & $6.9\pm0.1$ & $6.5\pm0.1$ & $6.8\pm0.1$ & $6.6\pm0.1$ & $6.72\pm0.07$ & $6.63\pm0.09$ & $6.77\pm0.07$ \\ [5pt]
      & $\mathrm \sigma_{\rm Fe}$ (keV) & $0.33^{+0.18}_{-0.18}$ & $0.4$ (fixed) & $0.39^{+0.18}_{-0.18}$ & $0.4$ (fixed) & $0.40^{+0.21}_{-0.21}$ & $0.4$ (fixed) & $0.43\pm0.17$ & $0.4$ (fixed) \\ [5pt]
      & $\rm {norm}_{\rm Fe}$ (photons cm$^{-2}$ s$^{-1}$) & $0.012\pm0.001$ & $0.011\pm0.001$ & $0.012\pm0.002$ & $0.012\pm0.001$ & $0.008\pm0.001$ & $0.006\pm0.001$ & $0.010\pm0.001$ & $0.009\pm0.001$ \\ [5pt]
      \code{gabs} & E$_{\rm CRSF}$ (keV) & $27.7\pm0.5$ & $28.8^{+1.0}_{-0.9}$ & $26.8\pm0.4$ & $27.3^{+0.8}_{-0.7}$ & $27.4^{+3.3}_{-1.6}$ & $30.0\pm1.7$ & $25.0^{+5.6}_{-2.9}$ & $30.0\pm1.7$ \\ [5pt]
      & $\mathrm \sigma_{CRSF}$ (keV) & 
      $7.6^{+0.9}_{-0.6}$ & $4.0^{+1.1}_{-0.8}$ & $6.7^{+0.6}_{-0.5}$ & $3.9^{+0.8}_{-0.7}$ & $6.09^{+3.06}_{-1.64}$ & $2.4\pm0.9$ & $4.87\pm1.95$ & $0.6\pm0.3$ \\ [5pt]
      & $\tau_{CRSF}$ & $0.64\pm0.16$ & $0.19\pm0.10$ & $0.64\pm0.14$ & $0.22\pm0.09$ & $0.29\pm0.19$ & $0.2\pm0.18$ & $0.11\pm0.11$ & $0.33\pm0.29$ \\ [5pt]

      \cline{1-10}
    & & & & & & & & & \\
      & $\chi^2$/dof & $33.79$/$36$ & $53/37$ & $36.29$/$36$ & $59.5/37$ & $31.65$/$33$ & $70.94/34$ & $28.62$/$34$ & $78.5/35$ \\ [5pt]
      
      & Flux$^b$(3--50~keV) & $9.70^{+0.01}_{-0.31}$ & $9.70^{+0.02}_{-0.03}$ & $8.96^{+0.02}_{-0.65}$ & $8.96^{+0.02}_{-0.02}$ & $4.16^{+0.05}_{-0.80}$ & $4.18^{+0.01}_{-0.01}$ & $4.52^{+0.01}_{-0.31}$ & $4.55^{+0.03}_{-0.01}$ \\ [5pt]
      \cline{1-10}

      \end{tabularx}

\end{sideways}
    \label{tab:phase_avg}
  
\end{table*}


\subsubsection{Pulse phase resolved spectroscopy}
\label{section:pulse_phase}

As discussed in the timing analysis, we observed a strong dependence of pulse profile on energy, which motivated us to further carry out phase resolved analysis of the source. We assigned phase boundaries and identified 6 sections based on various distinct features in the pulse profiles as shown in Figure \ref{fig:pulseprofiles}. We extracted the phase resolved spectra only for the high intensity state observations - Obs IDs A and B. This analysis was not carried out for the fainter states (Obs IDs C and D) due to poor photon count rates and limited statistics. In order to eliminate the excess contribution to the background from multiple anode layers, we use the top layer spectra of \code{LAXPC10} for phase resolved analysis. The source spectra were extracted using the standard pipeline \code{laxpc$\_$extract$\_$spectra}. The response files used were identical to that of the phase average analysis. The background spectra were generated using the procedure indicated by the faint source pipeline. 

\begin{figure*}
    \centering
     \includegraphics[scale=0.47]{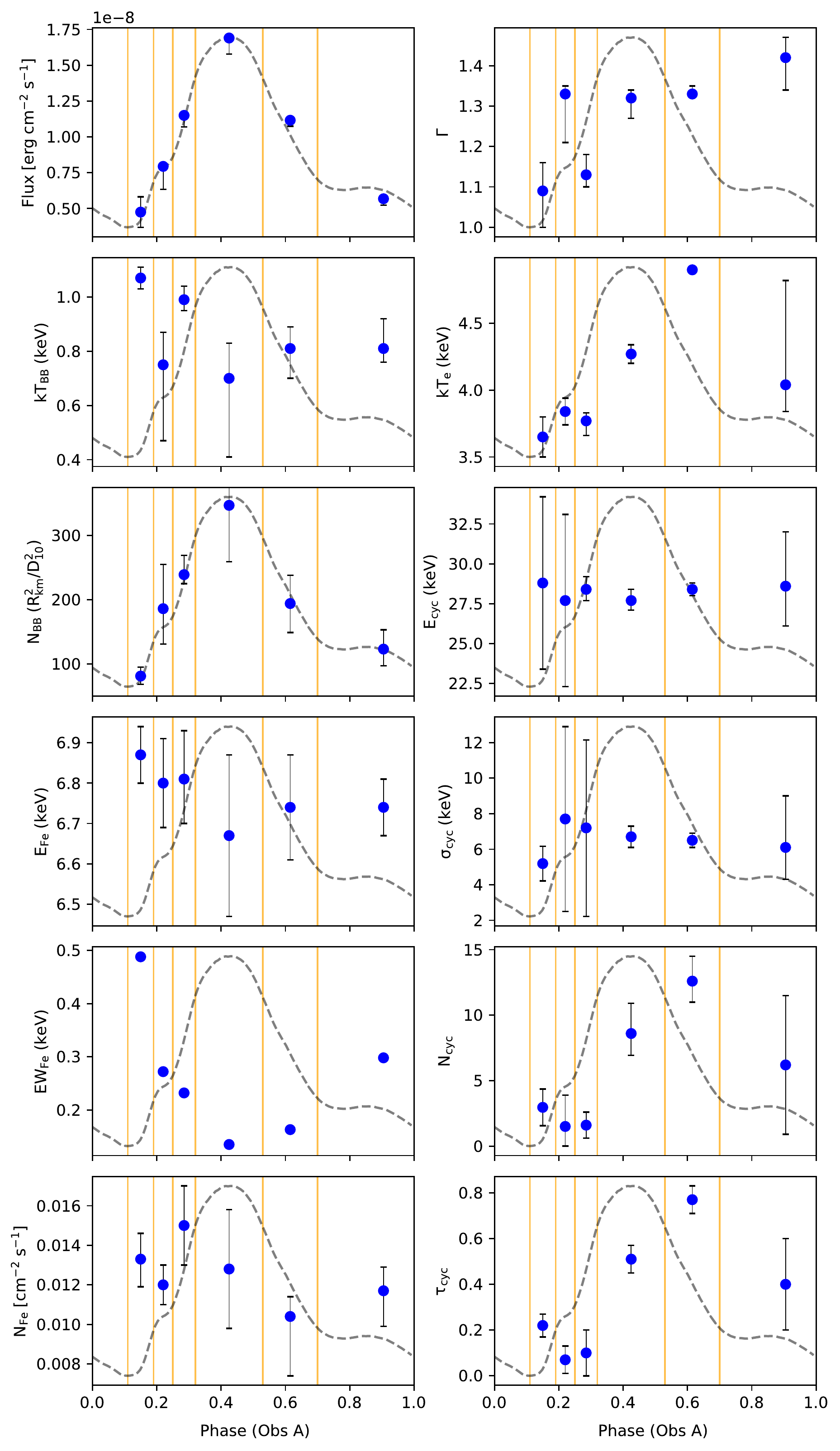}
    \includegraphics[scale=0.47]{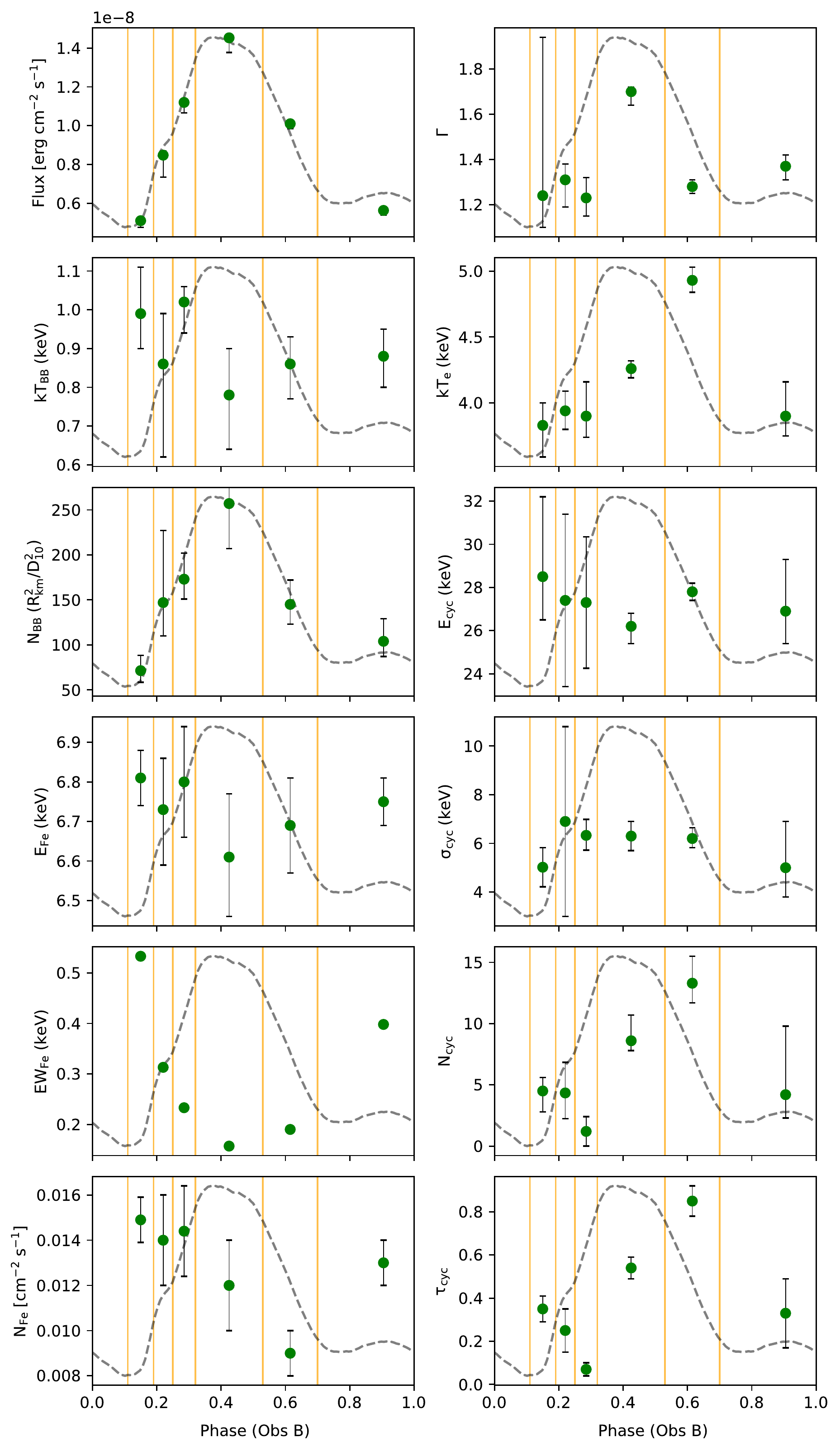}
    \caption{Best-fit parameters of pulse phase resolved spectroscopy are plotted as a function of pulse phase (Section \ref{section:pulse_phase}). The two panels on the left with blue data points correspond to Obs A and the two panels to the right with green data points represent Obs B. The pulse profile in the 3--80~keV range is shown as a grey dashed line for reference. The yellow vertical lines indicate the phase boundaries.}
    \label{fig:phase-res-par-var}
\end{figure*}

We restrict our spectral range to 50.0~keV, beyond which the spectrum has very limited S/N ratio. The spectra of all individual phases were fit with the same model used for phase averaged analysis - a comptonization model along with a blackbody emission (Model I). Fe line width were kept frozen to their respective phase averaged spectral fit values. We indicate the variation of the different spectral parameters for both Obs IDs (A \& B) as a function of pulse phase in Figure \ref{fig:phase-res-par-var}. The radius of the blackbody emitting region (indicated by the norm) varies in line with the pulse intensity as a function of pulse phase, indicating that during the peak pulse phase, a larger fraction of the blackbody emitting region becomes visible. The electron temperature (kT$_{e}$) responsible for comptonization shows an exceptionally large value at the declining phase. The other continuum parameters remain more or less uncorrelated with pulse phase. We also observe a variation of the CRSF optical depth $\big(\frac{1}{\sqrt{2\pi}}\frac{\rm depth}{\rm \sigma}\big)$ with pulse phase, with the declining phase showing the most depth compared to the rising phase. However, the Fe line strength varies inversely with pulse peak intensities, showing the least strengths at the peak phase. This picture of parameter variation and correlations is consistent across both Obs IDs.

\section{Discussion}

We have carried out a timing and spectral study of Cen X-3 in two luminosity states using \asat observations from 2016 and 2017. We demonstrate variations in the pulse profile from single to double peaked as the X-ray luminosity switched from a high to low state. Results from our spectral analysis indicate variation in continuum spectral parameters as well as the CRSF energy as a function of luminosity state. Further, we use measured  CRSF energy to obtain the NS surface magnetic field for further differentiating various QPO models. 

Cen X-3 is an interesting source, as it is the only known HMXB pulsar system that undergoes disk fed accretion \citep{raichur2009}; in fact, Cen X-3 is considered to be at the onset of its Roche Lobe overflow evolutionary stage. However, the reason for luminosity state changes in this system is still poorly constrained. A study using RXTE observations suggested that the different observed luminosity states in Cen X-3 could be due to the changing view of a warped accretion disk \citep{Raichur2008,Ferr2021}.

\subsection{Cyclotron line and QPO}
X-ray pulsars present the ideal setting to study effects of plasma in strongly magnetised environments. The energies of electrons traveling perpendicular to the magnetic field axis become quantized into Landau levels \citep{Meszaros1983}. The strongly peaked behavior of opacity at these discrete energy levels leads to a resonant scattering process. A distinct CRSF absorption profile is detected in X-ray spectra at energy values $E_{n} = n\frac{\hbar eB}{m_e}$ (assuming non-relativistic case), where $n$=1 corresponds to the transition between two consecutive levels. The detection of CRSF in X-ray pulsars is the most reliable method to constrain the magnetic field in X-ray pulsars. Using \asat results of spectral analysis of Cen X-3, we compute an average magnetic field of $\sim$2.3$\times$10$^{12}$ G. 

From a total of 36 X-ray pulsars with confirmed CRSF detections, about a third exhibit either a positive or negative correlation of the $E_{\rm CRSF}$ with luminosity (for details see Table 7 of \citealt{Staubert2019}). This prompted the development of an association between the mass accretion rate and the depth at which the in-falling material is decelerated \citep{Tsygankov2006,Mushtukov2015}. Above a certain `critical luminosity' (L$_{\rm crit}$), an anti-correlation between $E_{\rm CRSF}$ and $L_{\rm X}$ is observed (for example SMC X-2, V0332+53). In this regime, a radiatively-induced collisionless shock gets created at a certain height above the NS surface \citep{Basko-Suny1975} and therefore the CRSF forming region is higher up the in the accretion column, enabling better visibility. Several physical models attempt to explain the negative correlation at higher luminosities (for example, \citealt{Burnard1991} \citealt{Poutanen2013}, \citealt{Mihara2004}, \citealt{Nishimura2014}).  As the $\dot{M}_{\rm acc}$ increases, the radiative shock is formed higher up in the column, sampling lower magnetic field strengths and effectively leading to smaller measurements of $E_{\rm cyc}$. On the other hand, a positive correlation has been identified in low luminosity X-ray pulsars  like Vela X-1, A 0535+26 and GX 304-1 \citep{Staubert2007, Yamamoto2011, Klochkov2012}. This behavior has been explained by models that assume these sources had luminosities in the sub-critical regime \citep{Staubert2007}. Here, the plasma reaches very close to the NS surface. The CRSF is generated relatively deeper in the accretion column, which reflects a stronger B field and consequently a higher CRSF energy. 

The critical luminosity (L$_{\rm crit}$) serves as a boundary separating two regimes with opposite correlation patterns and several works have attempted to characterize this function  \citep{Basko-Suny1975,Becker2012,Mushtukov2015}. Above L$_{\rm crit}$, matter is deccelarated via radiative shocks from an accretion column \citep{Basko-Suny1975,Mushtukov2015} and below L$_{\rm crit}$, plasma decelerates via Coulomb collisions and reaches close to the NS surface \citep{ZelShakura1969}. Recent studies by \citet{Mushtukov2015} include a number of accretion flow geometry effects. They demonstrate that the L$_{\rm crit}$ is not a monotonic function of the magnetic field.  Their models are in good agreement with observed trends for some of the pulsars like V 0332+53 \citep{Doroshenko-2016} and more recently, for IGR J19294+1816 \citep{Raman2021}. 

\begin{figure}
    \centering
\includegraphics[scale=0.36,angle=0]{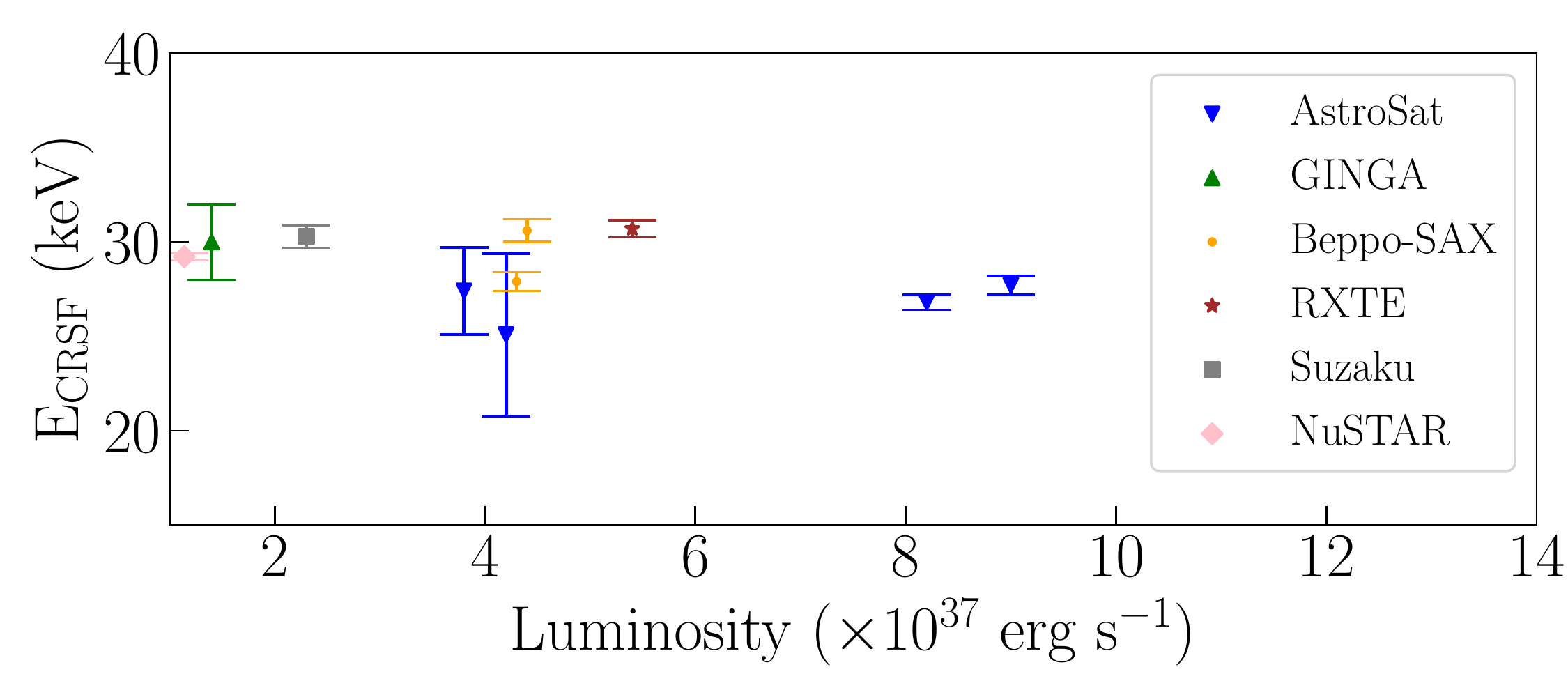}
    \caption{Figure shows all the past measurements of CRSF energies including the current \asat observations, as a function of luminosity state.}
    \label{fig:cyc-corr}
\end{figure}

In the context of Cen X-3, there have been reports of CRSF energies between 28-30~keV in different luminosity states observed using \textit{NuSTAR} and \textit{Suzaku} \citep{tomar2021}, GINGA \citep{Nagase1989}, Beppo-SAX \citep{Santangelo1998,Burderi2000}, RXTE \citep{Suchy2008}. Recent works using Suzaku and \textit{NuSTAR} observations \citep{tomar2021} and using \textit{Swift}-BAT \citep{Ji2019}, have compiled and compared the CRSF properties of Cen X-3 and have claimed that the CRSF energy in Cen X-3 has  remained steady at $\sim$31.6$\pm$0.2~keV for the last 14 years. Their work covered an X-ray luminosity range of 1.1-5.4$\times$10$^{37}$~ergs s$^{-1}$. Our current study of Cen X-3 using broadband \asat-LAXPC observations extends this study to higher luminosity states and shows that the line energy remains constant. In Figure \ref{fig:cyc-corr}, we indicate all the reported $E_{\rm CRSF}$ measurements.

CRSF properties for Cen X-3 have also been examined as a function of its pulse phase. In their work, \citet{Suchy2008} show that the CRSF energy follows the pulse profile shape with a variation ranging from 30--40~keV through an entire pulse period. A similar variation in CRSF centroid energy was reported using \bepposax data by \citet{Burderi2001MmSAI}. Our results indicate no such variation in the CRSF center and width parameters, except for the optical depth. The CRSF depth variation as a function of pulse phase has been observed many times in the past for Cen X-3 \citep{Suchy2008} as well as in other pulsars like 1A 1118-61 \citep{Maitra2012}. These optical depth variations can be attributed to the fact that the dipole geometry is being viewed from different angles at different pulse phases. Alternatively, this could also be suggestive of a more complex magnetic field structure in this source.

There are only a handful of HMXB pulsars that exhibit simultaneous CRSFs and QPOs, and Cen X-3 is one of them (see \citealt{Staubert2019} for details). The reason why these sources are crucial is that the presence of a CRSF helps to constrain the magnetic field strength and therefore, the magnetospheric radius. This would further be useful in obtaining estimates of predicted QPO frequencies that enable differentiation between various QPO models. QPOs are a common occurrence in X-ray pulsars and are understood to be related to the motion of inhomogenously distributed matter in regions close to the magnetospheric boundary and inner accretion disk \citep{PaulNaik2011}. QPO frequencies found in HMXB pulsars are typically in the range 1~mHz--40~Hz \citep{Psaltis2006}. Cen X-3 has demonstrated a wide range of QPO frequencies between 40-90~mHz, with no apparent correlation of the centroid frequencies or their RMS values with different luminosity states \citep{Takeshima1991,Raichur2008}. The $\sim$26~mHz QPO detected using \asat observations during the faint luminosity state is probably one of the slowest QPOs seen for this source. The measured source flux during the low luminosity state (Obs D) is 4.46$\times$10$^{-9}$~ergs/cm$^2$/s. Assuming a source distance of 8.7~kpc, we obtain a luminosity of 4.1$\times$10$^{37}$~erg/s. For a generic NS with mass 1.4~$M_{\odot}$ and a 10~km radius, the inner disk radius can be expressed as:
\begin{equation}
     r_{\rm M} = 3\times10^8   \rm{L}_{\rm{37}}^{-2/7} \rm{\mu}_{\rm 30}^{4/7}
\end{equation}
where L$_{37}$ is the luminosity in units of 10$^{37}$~ergs/s and $\mu_{30}$ is the magnetic moment in units of 10$^{30}$~G cm$^3$ \citep{FKR2002}. By using the measured value of B (2.3$\times$10$^{12}$~G) obtained from the detected CRSF, and assuming a 10~km radius NS, we infer a magnetic moment of $\mu_{30}$ of $\sim$2.3~G cm$^3$, which gives a magnetospheric radius of  $\sim$3.2$\times$10$^{8}$~cm. 

Depending upon the relationship between the NS spin frequency ($\nu_{\rm{spin}}$) and the QPO frequency ($\nu_{\rm{qpo}}$), several models have been traditionally invoked. The most common models include the Keplerian frequency model (KFM, \citealt{vanderklis1987}) and the magnetospheric beat frequency model (BFM, \citealt{AlparShaham1985}). The KFM predicts that the inhomogenous matter is modulated at the keplerian frequency corresponding to the inner accretion disk and therefore $\nu_{\rm{qpo}}=\nu_{\rm{keplerian}}$. This model is applicable only when $\nu_{\rm{qpo}}> \nu_{\rm{spin}}$ and is therefore not applicable to the 26~mHz QPO detected using current \asat observations. On the other hand, BFM predicts that the rotating inhomogeneity is further modulated by the rotating magnetic field and therefore shows up in the form of a QPO at the beat frequency between the $\nu_{\rm{keplerian}}$ at the inner disk and $\nu_{\rm{spin}}$. The inner accretion disk radius according to BFM is given as,

\begin{equation}
r_{\rm M,BFM} = \bigg(\frac{GM_{\rm NS}}{4 \pi^2 (\nu_{\rm spin}+\nu_{\rm qpo})^2}\bigg)^{1/3}    
\end{equation}

Using the r$_{\rm{M}}$ obtained from before (3.2$\times$10$^{8}$~cm), we can calculate the BFM-predicted QPO frequency which turns out to be $\sim$ 0.065~Hz, which is of a similar order, but higher, than the measured $\nu_{\rm qpo}$ of 0.026~Hz. 

Previously, the very early set of mHz QPOs from Cen X-3 were associated to modulations in the mass accretion rate on the NS polar caps \citep{Takeshima1991}. Later studies using RXTE by \citet{raichur2009} note that since the QPO frequency is uncorrelated to the luminosity state, it cannot be the result of material inhomogeneity at the inner accretion disk. Instead, they claim that the nearly constant QPO frequency range ($\sim$40~mHz) is primarily due to varying degree of obscuration by an aperiodically warped accretion disk \citep{Raichur2008, raichur2009}. This is one of the proposed explanations for the observed multiple flux states in Cen X-3 as well. 

A similar, more recent model that explains mHz QPOs in HMXB pulsars is the precessing warped accretion disk model by \citet{Shirakawa-Lai2002}. This model has been invoked to explain the mHz QPOs observed in 4U 0115+63 \citep{Jayashree2019}. They suggest that the observed QPO frequency is a reflection of a misaligned disk angular momentum vector and the magnetic field vector. The resulting precessional time scale is given by \begin{equation}
    \tau_{\rm prec} = 776\times \alpha^{0.85} \times L_{\rm 37}^{-0.71} \rm s
\end{equation}
where $\alpha$ is the accretion disk viscosity parameter (usually $<$1), and L$_{\rm 37}$ is the X-ray luminosity in units of 10$^{37}$ erg/s. Observational constraints on the $\alpha$ parameter typically lie in the rage 0.1--0.4, while numerical simulations predict a value that is smaller by a factor of 10 \citep{KingPrinLiv2007}. We assume an $\alpha$=0.02 based on the model fits carried out for another similar pulsar, 4U0115+63 \citep{Jayashree2019}. For the measured flux and corresponding luminosity in Obs D (L$_{37}\sim$1.7), we obtain a $\nu_{\rm{qpo}}\sim$52~mHz, which is slightly more than the \asat measured 26~mHz QPO. We figure that this model, like the BFM, is also capable of explaining the order of magnitude of the observed $\sim$mHz QPO seen in Cen X-3, although future observations would shed more light on the various QPO models. 

We also bring to note that not only is the QPO prominently detected solely during one of the low luminosity states, but its RMS exhibits a tentative increase near CRSF energies (see Section 3.1). In contrast to our observations, a previously reported 40~mHz QPO in Cen X-3 was found to be non-varying with respect to energy as well as uncorrelated to luminosity state \citep{Raichur2008}. All the QPO models discussed above suggest an inner-disk magnetospheric-boundary origin for the QPO; while the association to a CRSF opens up the possibility of the accretion column as possibly being another alternative region of origin for the QPO. Detailed exploration of such an association is out of scope for this work, but is highly recommended for future studies.


\subsection{Pulsed Fraction, spectral continuum parameters and X-ray luminosity state}
Distinct changes in the pulse profile for Cen X-3 observed in our \asat study allowed us to examine some of the fundamental correlations between the pulsed fraction (PF) in different energy bands and in various  luminosity states. The PF represents the relation between the emission from the accretion column (the pulsed component) and the emission from the other portions of the accretion flow (unpulsed component). It is computed by measuring the relative amplitude of the neutron star’s pulse profile. Several studies have been conducted to understand the trends observed in the pulse profiles and pulsed fractions in various energy bands and also in different luminosity states in HMXB pulsars (see for example, \citealt{Lut-Tsy2009}). For most sources, the pulsed fraction in general shows an increasing trend with energy, for example, GX 1+4 \citep{Ferrigno2007}, OAO 1657-415 \citep{Barnstedt2008}, EXO 2030+37 \citep{Klochkov2008}, SMC X-3 \citep{Zhao2018}, etc. Some other sources like 4U 0115+63 exhibit changes in pulse profile and corresponding PF close to the CRSF energy (and also near the higher harmonics, \citealt{Ferrigno2011}) owing to changes in accreting plasma flow properties and corresponding change in beam pattern of emission \citep{Tsygankov2006,Lut-Tsy2009}. In contrast, many other HMXB pulsars like A0535+262 and Cen X-3 \citep{Tsygankov2007}, etc. show no distinct features in the PF near the CRSF energies. As seen in the Figure \ref{fig:PFrac}, this is consistent from our current AstroSat analysis of Cen X-3 as well. \citet{Lut-Tsy2009} proposed a qualitative model (excluding gravitational light bending effects, etc.) to explain the variation of PF with energy. They propose that at higher energies, the contrast between minimum and maximum visible surfaces of the accretion column is highest, hence an increased PF.  

The behavior of PF as a function of luminosity state also shows interesting correlations (see \citealt{Lut-Tsy2009}, for a comprehensive study). For example, in Her X-1, the PF is fairly low during its low luminosity state. On the other hand, some pulsars like SXP 1323 \citep{Yang2018} and 2S1417-624 \citep{Gupta2019} exhibit an anti-correlation between the PF and L$_{\rm{X}}$ which is attributed to an increase in the unpulsed component or a decrease in the pulsed component, both of which can be attributed to contributions from additional modes of accretion that would eventually affect the beam configuration. Even more unusual and much harder to explain is the case of the pulsar V0332+52 that shows a positive PF-L$_{\rm{X}}$ correlation at low luminosities and a negative correlation at higher luminosities.

For a persistent pulsar like Cen X-3, which has exhibited a nearly constant E$_{\rm{CRSF}}$ over a wide range of fluxes in over three decades \citep{Nagase1989,Santangelo1998,Burderi2000,tomar2021}, it is interesting to see such a wide spread in the PF at lower energies ($<$20~keV) as compared to the energies above 20~keV. Based on \textit{INTEGRAL} observations, \citet{Lut-Tsy2009} suggested that any variation in the PF with L$_{\rm{X}}$ could be attributed to local inhomogeneities in the stellar wind or accretion flow. This picture is consistent with our claim that the accretion flow inhomogeneities are responsible for generating mHz QPOs as detected in one of low state \asat observations (Obs D). Detailed modeling of the energy dependent and asymmetric pulse profiles of Cen X-3 that were carried out on very early observations obtained from GINGA and OSO-8 have indicated the presence of two emission regions at the magnetic poles  \citep{Krauss1996}. It also suggested that the pulsed emission had contributions from a pencil beam as well as a fan beam, which could directly result in variations of the PF as seen by subsequent works including this paper.


In our current \asat analysis of Cen X-3 in two distinctly different luminosity states, we notice the following: i) as we move from the high to the low state, the pulse profile shape varies from single peaked to double peaked, ii) the pulsed fraction shows an increasing trend with energy although the soft band (3--10~keV) PF is higher by a factor of about 2 for the high state, and iii) the blackbody component shows a lower temperature (kT$\sim$0.7~keV) at the high intensity state, versus a kT$\sim$0.95~keV for the low intensity state;  a higher disk temperature indicates a larger disk contribution (unpulsed component) to the overall X-ray emission and naturally explains the lower PF in the low state. 

Using results obtained from this analysis we have observed that Obs A and Obs B have large detector count rates and broadband fluxes, making them brighter (or high) state observations, compared to the low/faint Obs IDs C and D. From the phase averaged spectral analysis, we further note that the spectral index is correlated with the source flux. The high state observations (Obs A and Obs B) have spectral indices $>$1.3 and $>$1.25, as seen using Model I and Model II, respectively; while that of the low states are ~1.1 and ~1.2, for the same two models. The spectrum therefore softens at higher flux states (See Figure \ref{fig:flux-vs-param}) as has been observed in other sources like EXO 2030+375 \citep{Wilson2008} and V0332+053 \citep{Mowlavi2006}. 
Using Model I, the brighter states also seem to prefer a slightly higher plasma temperature (kT$_e\sim$4.4 keV) compared to fainter states (with kT$_e\sim$3.8 keV). We also observe that the fainter states prefer higher cutoff energies and lower power law indices, compared to the brighter states, for the Model II. Such trends have also been observed in other pulsars such as 1A 0535+262, or GRO J1008-57, etc.  as shown in \citet{ReigNespoli2013}. This implies that, in the high state, the phase averaged spectral emission has a larger contribution from the soft photons from higher up in the accretion column, giving rise to a softer spectrum. On the other hand, during the low luminosity states, the radiation is emitted from a region that samples a larger portion of the scattered comptonized hard emission, resulting in an overall harder spectrum.\\
We also discuss some previous results that have explored spectral parameter variations across pulse phase. For \rxte observations, \citet{Suchy2008} model the continuum using a power law modified by a high energy cutoff at a source luminosity of $\sim$4$\times$10$^{37}$~erg~$s^{-1}$. Their results indicate dramatic variations in power law index as well as the cutoff and fold energy parameters as a function of pulse phase. The parameters are lower during the rise of the pulse profile and become prominent at the declining phase. This behavior is somewhat similar to what we found in \asat observations, in terms of the power law index and electron temperature at the declining pulse phase. We further note that the electron temperature seems to peak during the declining phase of the main pulse, while the spectrum hardens (see Figure \ref{fig:phase-res-par-var}) . We believe this declining flux could represent the pulse phase where the pulsed component is moving out of view. This could then expose the observer to the portion of the accretion column where hot electrons upscatter the seed photons into comptonized hard photons. This behavior is seen quite consistently in both the high state observations, Obs A and Obs B.\\
With these \asat observations, we note that the different luminosity states and the varying pulse phases together govern the changes in the spectral emission from different segments of the accretion column. The different luminosity states in Cen X-3 have been discussed in the literature to be due to varying levels of obscuration of the X-ray emitting region, although, varying mass accretion rates is not ruled out entirely \citep{Raichur2008,Ferr2021}. Since there exists a wealth of data for Cen X-3, future studies aimed at establishing a comprehensive theoretical framework will help understand disk accreting pulsars better.

\section{Summary and conclusions}

Since its discovery, Cen X-3 has been repeatedly examined using various high energy instruments. Here we summarize the key results and new insights that we have obtained using \asat observations:\\
i) Cen X-3 has exhibited the presence of a weak 26~mHz QPO at an rms of 3\%, at a luminosity of L$_{X}\sim$4$\times$10$^{37}$~ergs~s$^{-1}$. \\
ii) The detection of a CRSF has allowed us to measure the magnetic field to about 2.3$\times$10$^{12}$~G, consistent with previous measurements. However, since this observation also exhibits the presence of a mHz QPO, we utilize the B field measurement obtained using the CRSF to further examine various different QPO models. \\
iii) \asat sampled this source during two different luminosity states for which we have carried out pulse profile timing studies. The high luminosity states exhibit a broad single peaked profile, while the low states exhibit complex and double-peaking profiles, indicating changes in the beam pattern within a span of a few months.\\


\noindent\textbf{Data availability statement:}\\
The data underlying this article are available in the AstroSat data archive: https://astrobrowse.issdc.gov.in/astro$_:$\\archive/archive/Home.jsp \\


\textit{Acknowledgements} : The research is based on the results obtained from the \asat mission of the Indian Space Research Organisation (ISRO), archived at the Indian Space Science Data Centre (ISSDC). This work has also utilized the calibration data-bases and auxillary analysis tools developed, maintained and distributed by \asat-SXT team with members from various institutions in India and abroad.

\bibliography{bibtex}{}
\bibliographystyle{mn2e}

\end{document}